\documentclass[12pt]{article}

\usepackage{graphicx}

\usepackage{amssymb,amsfonts,amsmath}

\begin{document}

\begin{center}
\begin{Large}
{\bf Multifractal Network Generator}
\end{Large}

\medskip

\begin{large}
Gergely Palla$^{\dagger}$, L\'aszl\'o Lov\'asz$^{\nmid}$ and Tam\'as Vicsek$^{\dagger\ddagger}$
\end{large}
\end{center}

\bigskip

\noindent $^{\dagger}$Statistical and Biological Physics Research Group of
HAS, E\"otv\"os University, Budapest, Hungary,

\noindent $^{\nmid}$Institute of Mathematics, E\"otv\"os University, Budapest, Hungary,

\noindent $^{\ddagger}$Dept. of Biological Physics, E\"otv\"os University,
 Budapest, Hungary.

%\author{Gergely Palla\affil{1}{Statistical and Biological Physics Research Group of HAS, E\"otv\"os University, Budapest, Hungary},
%L\'aszl\'o Lov\'asz\affil{2}{Institute of Mathematics, E\"otv\"os University, Budapest, Hungary},
%\and
%Tam\'as Vicsek\affil{1}{Statistical and Biological Research Group of HAS, E\"otv\"os University, Budapest, Hungary}\affil{3}{Dept. of Biological Physics, E\"otv\"os University, Budapest, Hungary}}

%\contributor{Submitted to Proceedings of the National Academy of Sciences
%of the United States of America}

%\maketitle

%\begin{article}
\begin{abstract}
We introduce a new approach to constructing networks with realistic features.
Our method, in spite of its conceptual simplicity (it has only two parameters)
is capable of generating a wide variety of network types with prescribed
statistical properties, e.g., with degree- or clustering coefficient
distributions of various, very different forms. In turn, these graphs
can be used to test hypotheses, or, as models of actual data. The method
is based on a mapping between suitably chosen singular measures defined
on the unit square and sparse infinite networks. Such a mapping has the
 great potential of allowing for graph theoretical results for a variety of
 network topologies. The main idea of our approach is to go to the infinite
 limit of the singular measure and the size of the corresponding graph
simultaneously. A very unique feature of this construction is that the
complexity of the generated network is increasing with the size.
We present analytic expressions derived from the
parameters of the -- to be iterated-- initial generating measure for
such major characteristics
of graphs as their degree, clustering coefficient and assortativity coefficient
distributions. The optimal parameters of the generating measure are
 determined from a simple simulated annealing process.
Thus, the present work provides a tool for researchers from a variety
of fields (such as biology, computer science, biology, or complex
systems) enabling them to create a versatile model of their network
data.
\end{abstract}

%\keywords{multifractal| random graph| graph sequences}

%\abbreviations{SAM, self-assembled monolayer; OTS,
%octadecyltrichlorosilane}

\section{Introduction}
As our methods of studying the features of our environment are
 becoming more and more sophisticated, we also learn to appreciate the
complexity of the world surrounding us. The corresponding systems
 (including natural, social and technological phenomena) are made of
 many units each having an important role from the suitable functioning
 of the whole. An increasingly popular way of grabbing the intricate
 structure behind such complex systems is a network or graph
 representation in which the nodes correspond to the units
 and the edges to the connections between the units of the original system
 \cite{albert-revmod,dm-book,newman-book}.
It has turned out that networks corresponding to realistic systems can be
highly non-trivial, characterized by  a low average distance
combined with a high average clustering coefficient \cite{watts-strogatz},
anomalous degree distributions \cite{F3,barabasi-albert} and an intricate
modular structure \cite{gn-pnas,our-nature,fortunato-new}.
A better understanding of these graphs is expected and,
in many cases have been shown, to be efficient in designing and controlling
complex systems ranging from power lines to disease networks
\cite{laci-disease}.

As increasingly complex graphs are considered a need for a better
representation of the graphs themselves has arisen as well. Sophisticated
visualization techniques emerged \cite{drawing_contest} and a
series of new parameters have been introduced
over the years \cite{albert-revmod,dm-book,newman-book}.
Very recently one of us
(L.L.) proved that in the infinite network size limit, a
 dense graph's adjacency matrix can be well represented by a continuous
function $W(x,y)$ on the unit square \cite{lovasz-szegedy,lovasz_cikk}. 
A similar
approach was introduced by Bollob\'as et al.
\cite{Bollobas-cikk,Bollobas_chapter} and
used to obtain convergence and phase transition results for
inhomogeneous random (including sparse) graphs. This two variables
symmetric function
 (which can have a very simple form for a variety of interesting
  graphs, and was supposed to be either continuous or almost everywhere continuous) predicts the probability whether two nodes are
 connected or not. (The non-trivial relations between the limiting objects 
of graph sequences and 2d functions are discussed in more details 
in the Supporting Information). 
 In this paper we develop the above ideas further
 in order to obtain simple and analytically treatable models of
 random graphs with a level of complexity growing together with
 their size. This is an important conceptual step acknowledging a
 rather natural expectation: the internal organization of larger
 networks is more complex than those of the smallest ones. (E.g.,
 the social contacts in large universities are much more
 structured than in an elementary school, which is in part due to
 the underlying hierarchical organization of almost every large
 networks we know of.)

In a sense, using a function to represent a network is very much like
using a model to describe a network. Models in the context of networks
have been playing a crucial role since they are ideal from the
 point of grabbing the simplest aspects of complex structures
and thus, are extremely useful in understanding the underlying principles.
Models are also very useful from the point of testing hypotheses about
measured data. Indeed, many important and successful models have been
 proposed over the past 10 years to interpret the various aspects
of real world networks. However, a considerable limitation of these
models is that they typically explain a particular aspect of the network
 (clustering, a given degree distribution, etc), and for each
new --to be explained-- feature a new model had to be constructed.

In the recent years, generating graphs with desired properties has
attracted great interest. A few
remarkable methods have been proposed, including the systematic approach
for analyzing network topologies by Mahadevan et al., using
the $dK$-series of probability distributions \cite{dk-series}.
These distributions specify all
degree correlations within $d$ sized subgraphs of a given graph,
with $0K$ reproducing the average degree, $1K$ the degree distribution,
$2K$ the joint degree
 distribution, etc. Several methods for generating
random graphs having a predefined finite $dK$-series
were also given in \cite{dk-series}, (with typically $d\leq3$).
Most important of these techniques
is based on rewiring of the links, as this turned out to be the only
efficient tool in practice.

The concept of characterizing a network
via the frequencies of given sub-graphs (forming a series with increasing
size) is at the heart of the exponential random graph model as well
 \cite{Frank-exp,Wasserman-exp,snijders-exp}.
In this approach a possible sub-graph $g$ (e.g., a pair of connected
 nodes, a wedge of a pair of links sharing a node, a triangle,
 etc.) is assigned a parameter $\eta_g$ related to the frequency
of the sub-graph,
 and the probability of a given network configuration is assumed
to be proportional to $\exp(\sum \eta_g  n_g)$, where $n_g$ denotes
 the number of sub-graphs occurring in the network. The $\eta$
parameters for a studied network are usually estimated using maximal
likelihood techniques.

The $dK$-series method and the exponential random graph model can be
viewed as bottom-up approaches: in the first order approximation of
the studied network we concentrate on the frequency of the most
simple object (an edge), when this is reproduced correctly we move on
 to a slightly more complex sub-graph and so on.
The series of sub-graphs from small/simple to large/complex are ordered
into a sort of hierarchy.
However, in a realistic scenario we stop in the above process at a relatively
early stage, since on
one hand most important properties of the networks are usually reproduced
already, on the other hand including ``higher order'' sub-graphs becomes
 computationally very expensive.

Hierarchy, self-similarity and fractality are very important concepts when
describing complex systems in nature and society, and turned out to be
relevant in network theory as well \cite{Tamas_fract,Havlin_fract,Newman_hier}.
 Very recently two important network
models have been introduced which are intrinsicly hierarchical, yet
show general features. Avetisov et al. proposed in \cite{Avetisov} the
construction of random graphs having an adjacency matrix equivalent
 to a $p$-adic randomized locally constant Parisi matrix, one of the
 key objects in the theory of spin glasses \cite{Parisi_matrix}. This symmetric
matrix has a hierarchic structure, and its elements are Bernoulli distributed
random variables (taking the value of 1 with probability $q_{\gamma}$ and
 the value 0 with probability $1-q_{\gamma}$, where $\gamma$ counts the
hierarchy levels). An interesting feature of this construction is that
any sub-graph belonging to a specific hierarchy level $\gamma$ is equivalent
 to an Erd\"os-R\'enyi random graph \cite{E-R},
nevertheless the overall degree distribution can be scale-free.

The Kronecker-graph approach introduced by Leskovec et al. is centered
around hierarchic adjacency matrices as well, however in this case
the self similar structure is achieved by Kronecker multiplication as
 follows \cite{Leskovec_elso}. Starting
 from a small adjacency matrix $A^1$, (where $A^1_{ij}=1$
if nodes $i$ and $j$
 are linked, otherwise $A^{1}_{ij}=0$), at every iteration we replace each
current matrix element by $A^1$ multiplied by the matrix element itself, hence
 enlarging the matrix by a factor given by the size of $A^1$. In the
stochastic version of this model the elements of $A^1$ are replaced by
real numbers between 0 and 1, and at the final stage of the multiplication
process we draw a link for each pair of nodes with a probability given
by corresponding element in the obtained stochastic adjacency matrix.
According to the results, the Kronecker-graphs obtained in this approach
can mimic several properties of real networks (heavy tails in the degree
 distribution, and in the eigenvalue spectra, small diameter,
densification power law) simultaneously. Furthermore, in \cite{Leskovec_two}
Leskovec and Faloutsos presented a scalable method for fitting real networks
with Kronecker graphs.

We note that link probability matrices similar to the previous examples
can be also used for community detection as pointed out by Nepusz et al.
in \cite{Nepusz_chapter,Nepusz_cikk}. In their approach (inspired
 by Szemer\'edi's regularity lemma \cite{Szemeredy}) the diagonal
elements of the matrix give the the link density inside the corresponding
 communities, whereas the off-diagonal elements correspond to the
link probabilities between the groups.

 In summary, a plausible classification of the emerging graph generating 
procedures/approaches involves the following types. Generating
graphs as {\bf i)} stochastic growth processes (e.g.,
\cite{albert-revmod}), {\bf ii)} as a process of connecting or rewiring
nodes according to prescribed probabilities (\cite{watts-strogatz,Molloy_Reed_95,Molloy_Reed_98,Chung_lu,Boguna}), 
{\bf iii)} accepting varying configurations with a
prescribed probability \cite{dk-series,Frank-exp,Wasserman-exp,snijders-exp}, 
{\bf iv)} by
deterministically or stochastically obtaining its adjacency matrix
from simpler initial matrix, \cite{Avetisov},
\cite{Leskovec_elso},\cite{Leskovec_two},  {\bf v)} from a function $W(x,y)$ on
the unit square providing a value for the probabilities of node pair
connections \cite{lovasz-szegedy,lovasz_cikk,Bollobas-cikk}.

Rewiring and the related construction techniques do not provide a
clue how a complex network emerges from a simple rule. On the other
hand, generating a graph from a fixed function/measure does not
result in networks with increasing complexity.
Our approach can be considered as a combination of iv) and v) (thus,
combining their advantages), assuming that in the infinitely large
network limit the right representation is a singular measure
(nowhere continuous function).

Thus, here we introduce a new method to constructing random graphs
inheriting features from real networks.  The main idea of our
approach is to replace $W(x,y)$ by a fractal (singular) measure
(also called multifractal),
 and go to the limit of infinitely fine resolution of the measure and
the infinitely large size of the generated graph simultaneously.
Consequently, the complexity of the obtained network is increasing with
the size. Another advantage of this approach is that the statistical
features characterizing the network topology, e.g., the degree
 distribution, clustering coefficient, degree correlations, etc, can
be simply calculated analytically. For generating networks with a
given prescribed statistical feature (e.g., a given degree
distribution), the optimal parameters of the generating measure
defining the multifractal can be determined from a simple simulated
annealing process.

\section{Model}
\begin{figure}
\begin{center}
\centerline{\includegraphics[width=\textwidth]{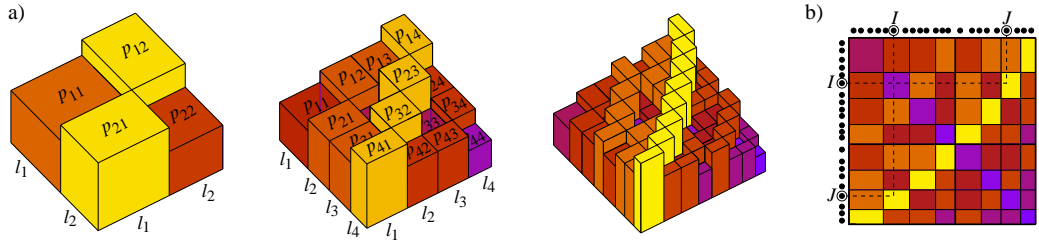}}
\caption{Schematic illustration of the multifractal graph generator. 
a) The construction of the link probability measure. 
We start from a symmetric generating measure on the unit square 
defined by a set of probabilities $p_{ij}=p_{ji}$ associated 
to $m\times m$ rectangles (shown on the left). 
In the example shown here $m=2$, 
the length of the intervals 
defining the rectangles is given by $l_1$ and $l_2$ respectively, and the
 magnitude of the of the probabilities is indicated by both the height and 
the color of the corresponding boxes. 
The generating measure is iterated by recursively 
multiplying each box with the
generating measure  itself as shown in the middle and on the right,
yielding $m^k\times m^k$ boxes at iteration $k$. The variance of the 
height of the boxes (corresponding to the probabilities associated 
to the rectangles) becomes larger at each step, producing a surface which is
getting rougher and rougher, meanwhile the symmetry and the 
self similar nature of the multifractal is preserved. 
b) Drawing linking probabilities from the obtained measure. We assign
random coordinates in the unit interval to the nodes in the graph,
and link each node pair $I,J$ with a probability given by the probability
 measure at the corresponding coordinates.}
\label{fig:magyaraz}
\end{center}
\end{figure}

\begin{figure}
\begin{center}
\centerline{\includegraphics[width=\textwidth]{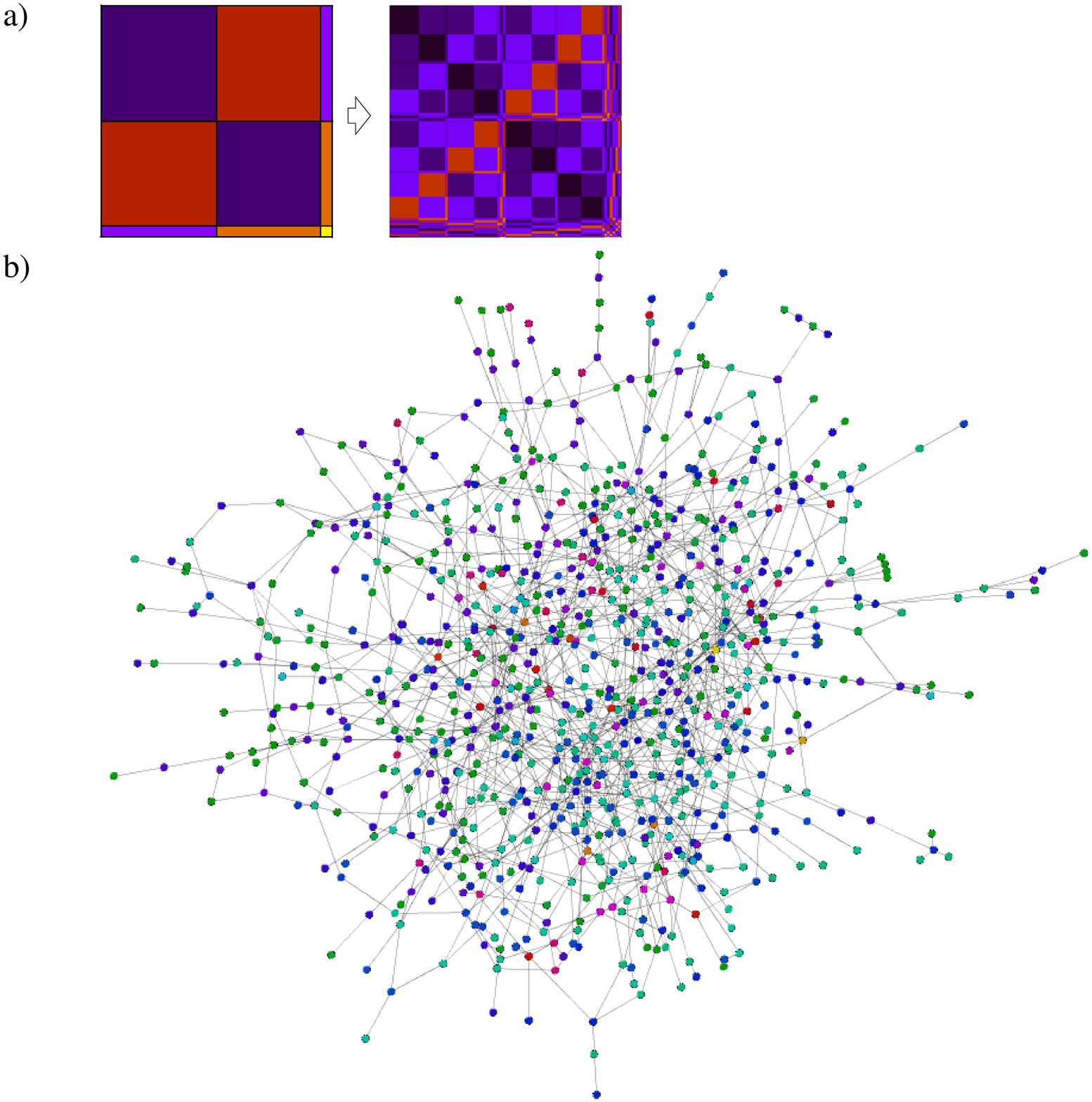}}
\caption{A small network generated with the multifractal network generator. 
a) The generating measure (on the left) and the link probability measure 
(on the right).
 The generating measure consists of 3$\times$3 
rectangles for which the magnitude of the associated probabilities is indicated
by the color in a similar fashion to Fig.\ref{fig:magyaraz}. 
The number of iterations, $k$, is set to $k=3$, thus the final 
link probability measure consists of 27$\times$27 
boxes, as shown in the right panel. (Note that $k=1$ corresponds to 
the generating measure in our convention). 
b) A network with 500 nodes generated from 
the link probability measure. The colors of the nodes were chosen as 
follows. Each row in the final linking probability measure was assigned a 
different color, and the nodes were colored according to their position 
in the link probability measure. (Thus, nodes 
falling into the same row have the same color). }
\label{fig:small_example}
\end{center}
\end{figure}

The network generation has three main stages in our approach: we
start by defining a generating measure on the unit square, next
we transform the generating measure through a couple of iterations
into a link probability measure, and finally, we draw links between 
the nodes using the link probability measure. The generating measure 
is defined as follows. 
We identically divide both $x$ and $y$ axis of the unit square
to $m$ (not necessarily equal) intervals, splitting it to $m^2$
rectangles, and assign a probability $p_{ij}$ to each rectangle
($i,j\in[1,m]$ denote the row and column indices). The probabilities
must be normalized, $\sum p_{ij}=1$ and symmetric $p_{ij}=p_{ji}$.
Next, the link probability measure is obtained by recursively multiplying
 each rectangle with the generating measure $k$ times 
(which is equivalent to taking the $k$-th tensorial product of the 
generating measure).
 This results in
$m^{2k}$ rectangles, each associated with a linking probability
 $p_{ij}(k)$ equivalent to a product of $k$ factors from the
original generating $p_{ij}$. In our convention $k=1$ stands for
the generating measure, thus, a link probability measure at $k=1$ is equivalent
 to the generating measure itself. 
Finally, we distribute $N$ points independently,
 uniformly at random on the $[0,1]$ interval, and link each pair
with a probability given by the $p_{ij}(k)$ at the given coordinates.
The above process of network generation is illustrated in 
Fig.\ref{fig:magyaraz}, whereas in Fig.\ref{fig:small_example}. we show
a small network obtained with this method.

We note that our construction could be made even more general by replacing
the ``standard'' multifractal with the $k$-th tensorial product of a 
symmetric 2d function $0\leq W(x,y)\leq 1$ defined on the unit square. 
Although the resulting 
$W_k(x_1,...,x_k,y_1,...,y_k)=W(x_1,y_1)\cdots W(x_k,y_k)$ 
function is $[0,1]^{2k}\rightarrow[0,1]$ instead of $[0,1]^2\rightarrow[0,1]$,
with the help of a measure preserving bijection between $[0,1]$ and
 $[0,1]^{k}$ it could be used to generate random graphs in the same 
manner as with our multifractal.

The diversity of the linking probabilities $p_{ij}(k)$ (and correspondingly,
the complexity of the generated graph) is increasing with the number
of iterations, just like in case of a standard multifractal.
In order to keep the generated networks sparse, we must ensure that
the average degree, $\left< d\right>$ of the nodes does not change
between subsequent iterations.
This can be achieved by an appropriate choice of the number of nodes
as a function of $k$, using the following relation:
\begin{equation}
\left< d\right>=N\sum_{i=1}^{m^k}\sum_{j=1}^{m^k}p_{ij}(k)a_{ij}(k),
\end{equation}
where $a_{ij}(k)$ denotes the area of the box $i,j$ at iteration $k$.
In the special case of equal sized boxes $a_{ij}(k)=m^{-2k}$, and due to
the normalization of the linking probabilities the above expression simplifies
to $\left< d\right>=Nm^{-2k}$. Thus, to keep the average degree constant
when increasing the number of iterations for a given generating measure, the
 number of nodes have to be increased exponentially with $k$.

\section{Statistical Methods}

One of the main advantages of our model is that the statistical
properties characterizing the network topology can be calculated
analytically. An important observation concerning our model is that
nodes having coordinates falling into the same row (column) of the
link probability measure are statistically identical. This means
that e.g., the expected degree or clustering coefficient of the
nodes in a given row is the same. Consequently, the distributions
related to the topology are composed of sub-distributions associated
with the individual rows.

Let us concentrate on the degree distribution first, which can be
expressed as
\begin{equation}
\rho^{(k)}(d)=\sum_{i=1}^{m^k}\rho_i^{(k)}(d)l_i(k),
\end{equation}
where $\rho_i^{(k)}(d)$ denotes the sub-distribution of the nodes in
row $i$, and $l_i(k)$ corresponds to the width of the row
(giving the ratio of nodes in row $i$ compared to the number of
total nodes). These $\rho_i^{(k)}$ can be calculated using the generating
function formalism as shown in the Appendix, resulting in
\begin{equation}
\rho_i^{(k)}(d)=\frac{\left< d_i(k)\right>^{d}}{d!}e^{-\left< d_i(k)\right>},
\label{eq:row_deg_dist}
\end{equation}
where $\left< d_i(k)\right>=N\sum_j p_{ij}(k)l_j(k)$ denotes the average degree
of nodes in row $i$.
 Even though the degree distribution of
nodes in a given row follows a Poisson-distribution according to
(\ref{eq:row_deg_dist}), the overall degree distribution of the generated
 graph can show non-trivial features, as will be demonstrated later.

\begin{figure}[h!]
\centerline{\includegraphics[width=\textwidth]{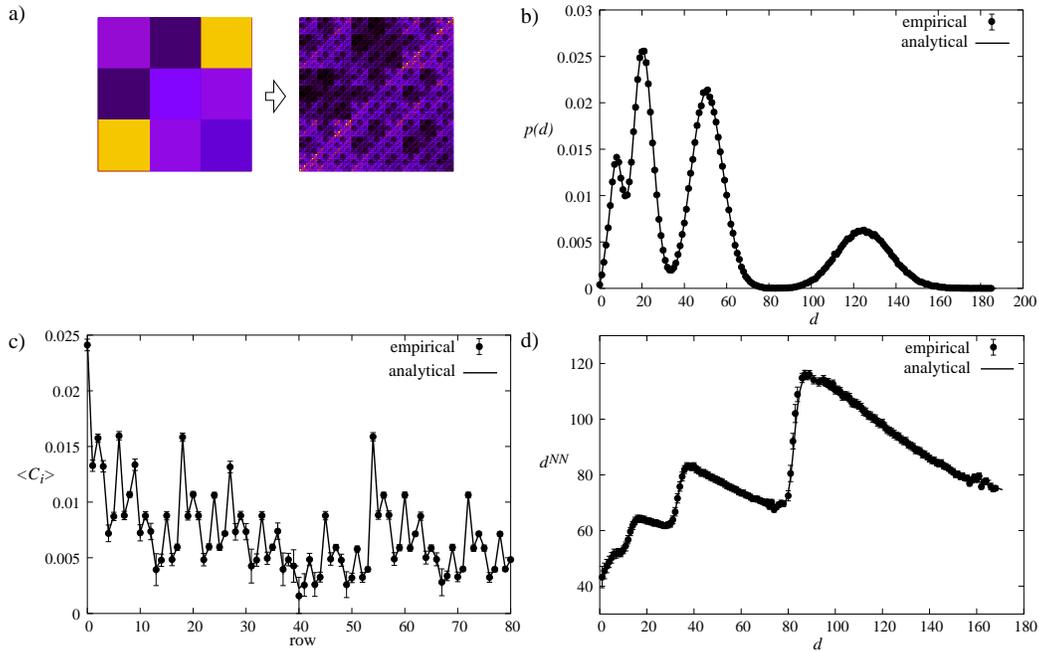}}
\caption{ Comparison between the analytical and empirical result for
a randomly chosen generating measure. 
a) The generating measure (left) and the link probability measure (right). 
The number of rows in the generating measure was set to
 $m=3$ with equal box lengths $l_i=1/3$, the corresponding
 initial linking probabilities $p_{ij}$ were chosen randomly. 
The number of iterations
 was set to $k=4$, resulting in the linking probability measure shown on
 the right. This link probability measure was used to generated 
100 samples of random networks with $N=5000$ nodes each. 
b) The degree distribution obtained by averaging over the samples 
(symbols), plotted together with the analytical result 
obtained from eq.(\ref{eq:row_deg_dist}) (continuous line), showing 
very good agreement. The error-bars (showing the standard error of the mean)
are smaller than the symbols.
The average clustering coefficient $\left< C\right>$ of nodes 
falling into the same row of the final link probability measure 
and the nearest neighbors average degree as a function of the node degree
are plotted in panels (c) and (d) respectively.
}
\label{fig:calc_and_emp}
\end{figure}

Similarly to the degree distribution, the clustering coefficient and
 the average nearest neighbors degree can be calculated analytically
as well in a rather simple way (as given in the
Supporting Information). According to Fig.\ref{fig:calc_and_emp}b-d, 
the analytical results for the quantities above
 are in very good agreement with the empirical distributions,
 (obtained by generating a number of sample graphs for
the chosen parameters).
The use of analytic formulas
instead of empirical distributions can significantly speed up the 
optimization of the
generating measure with respect to some prescribed target property.

\section{Results}

Depending on the choice of the generating measure and the box boundaries,
our method is capable of producing graphs with diverse properties. However,
 to generate a random graph with prescribed features in our approach we need
to optimize the generating measure with respect to the given
requirements. Let us suppose that the number of nodes in the graph
to be generated is given. In this case we have two parameters: the
number of boxes in the generating measure (given by $m^2$), and the
number of iterations, $k$. The actual $p_{ij}$ and box boundaries
are ``self adjusting'', as we shall describe in the following.

Let us denote the property to which we are optimizing the generating measure
by ${\cal F}$.  A conceptually simple example is when our goal is to obtain a
network with a given degree distribution, in this case ${\cal F}$ is
equivalent to $p(d)$. In principle, ${\cal F}$ depends on $p_{ij}$, $l_i$, $k$,
 $N$ (and in an implicit way on $m$, through the box sizes and
linking probabilities). However, as $m,k$ and $N$ are kept constant, we
 discard them from the notation and write the ``value'' of the property
corresponding to a given choice of $p_{ij}$ and $l_i$ as ${\cal F}(p_{ij},l_i)$.
(Note that in most cases ${\cal F}(p_{ij},l_i)$ is actually a high dimensional
object, e.g, a degree distribution, and not a real number). The target
value of the property to which we would like the system to converge is denoted
 by ${\cal F}^*$.

In order to be able to make the studied property of the generated
network converge to the goal ${\cal F}^*$, we have to define a way
to judge the quality of the actual ${\cal F}(p_{ij},l_i)$. In other
words, we have to define a sort of distance or similarity between
${\cal F}(p_{ij},l_i)$ and ${\cal F}^*$. This distance/similarity
measure can be used as an {\it energy function} during a so called 
simulated annealing procedure, and we shall denote it by $E[{\cal
F}(p_{ij},l_i),{\cal F}^*]$. The actual form of this function
depends on the actual choice of the property, e.g., in case of
optimizing the degree distribution a plausible choice is the sum of
the relative differences between the degree distributions:
\begin{equation}
E[{\cal F}(p_{ij},l_i),{\cal F}^*]=-\sum_{d}\frac{\left|\rho^{(k)}(d)-\rho^*(d)\right|}
{\max(\rho(d),\rho^*(d))},
\end{equation}
where $d$ runs over the degrees, $\rho^{(k)}(d)$ is the value of the actual
 degree distribution at degree $d$ and $\rho^*(d)$ is the value of the target
 degree distribution at the same degree.

In the simulating annealing we also define a temperature, $T$, which
is decreased slowly during the process. The process itself consist of many
Monte-Carlo steps, and in one step we try to change one of the linking
probabilities or one of the box boundaries by a small amount, 
following the Metropolis algorithm \cite{Metropolis}.
If the energy $E_2$ after the change is smaller than the energy $E_1$ before,
 the change is accepted. In the opposite case, the change is
 accepted by a probability given by $P=\exp[-(E_2-E_1)/T]$.

\begin{figure}[h!]
\centerline{\includegraphics[width=\textwidth]{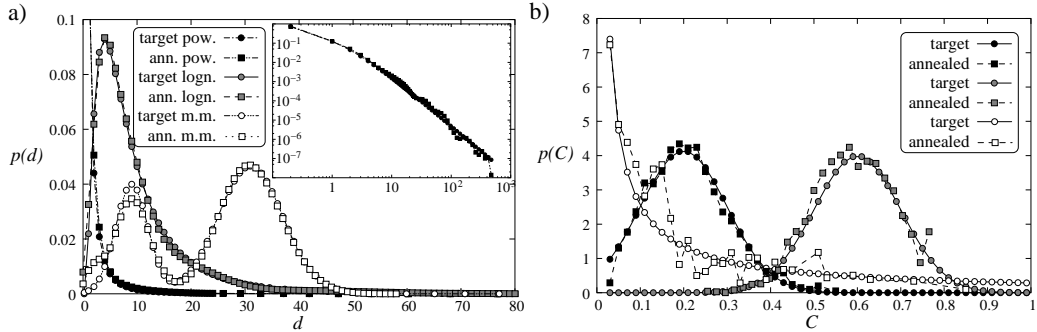}}
\caption{
Optimizing the generating measure with respect to different target properties.
During the optimization process the number of nodes, $N$, the
 number of rows in the generating measure, $m$, and the number of iterations,
 $k$, are kept constant, only the probabilities $p_{ij}$ and the length of
 the intervals $l_i$ defining the generating measure are adjusted. 
The typical value of the constant parameters in our experiments were
$N=10000-20000$, $m=3-4$ and  $k=3-5$.
a) Optimizing with respect to different degree distributions. 
 The target distributions are shown with circles, whereas the
 corresponding results at the end of the optimization procedure are
 marked by squares. The black symbols come from an experiment where the target
 was a power-law degree distribution (the inset shows this on log-scale), 
the gray symbols correspond
 to a setting with a log-normal target, whereas the white symbols show the
 results of an experiment with a bi-modal target distribution. 
b) Optimizing with respect to different
clustering coefficient distributions.}
\label{fig:anneal_ddists}
\end{figure}

The above procedure can be generalized in principle to optimizing with
 respect to multiple properties simultaneously as well. However, for
 simplicity here we consider the optimization of the different properties
 separately. In Fig.\ref{fig:anneal_ddists}a 
we show the results for optimizing the
 generating measure with respect to various target degree distributions.
Although the three chosen targets are rather different,
(a scale-free distribution, a log-normal one, and a bi-modal distribution),
our method succeeded in finding a setting of $p_{ij}$ and $l_i$ producing
a degree distribution sufficiently close to the target. Similarly,
 in Fig.\ref{fig:anneal_ddists}b the results from optimizing with
 respect to three clustering coefficient distributions are displayed showing
 again a reasonable agreement between the targets and the results.

\section{Discussion}

Our approach raises a number of fundamental graph theoretical and practical
questions. Should we expect
that large real graphs converge to some limiting network in a strict
sense of the convergence? Or, alternatively, their structure cannot
be mapped onto a fixed function, and only an ever changing (with the
size of the network) measure (in the infinite network size limit
becoming singular) can be used to reflect the underlying structural
complexity? This would be in contrast with the consequences of the
renowned Szemer\'edi Lemma \cite{Szemeredy} valid for arbitrary dense graphs.

Although it can be shown analytically (see SI) that in the infinitely
large network size limit our construction converges to a relatively
simple graph, the convergence to this structure is extremely slow.
According to our numerical studies, there is a very extensive region
between the small and infinite regimes in which a well defined,
increasingly complex structure emerges as our method is applied.
Details about aspects of the slowness of convergence involving an
extremely slow growth of the relative number of isolated nodes and the
appearance of oscillations are given in the SI.

In summary, our results demonstrate that it is possible to use simple models to
construct large graphs with arbitrary distributions of their
essential characteristics, such as degree distribution, clustering
coefficient distribution or assortativity. In turn, these graphs
can be used to test hypotheses, or, as models of actual data.
The combination of the
tensorial product of a simple generating measure and simulated
annealing technique leads to small (in practice 3x3 to 5x5) matrices
representing the most relevant statistical features of observed
networks. A very unique feature of this construction is that the
complexity of the generated network is increasing with the size.
In addition, the multifractal measure we propose is likely to result
in networks displaying aspects of self-similarity in the spirit of the
related findings by Song et al. \cite{Havlin_fract}.

\section*{Acknowledgments}
We thank fruitful discussions with G\'abor Elek.
This work was partially supported by the the Hungarian National Science
Fund (OTKA K68669, K75334 and T049674), the National Research
 and Technological Office (NKTH, Textrend) and the J\'anos
Bolyai Research Scholarship of the Hungarian Academy of Sciences.

%\end{article}

\section*{Appendix}

\subsection*{A1 Thermodynamic limit of graph sequences}

The infinitely large system size limit is very important in statistical
 physics, as the properties of the studied phenomena are manifested in
 the most pure way in this limit. Similarly, in complex network theory
 the infinitely large limit graph  of a converging graph sequence
can be considered as a ``platonic'' network exhibiting the fundamental
common properties of the graphs in the sequence in the most pure way.
But under what conditions can we say that a given sequence of graphs
is converging to something non-trivial and yet sufficiently universal
 to be conceptually meaningful? A simple intuitive condition is that
the statistical features commonly used to characterize a network
 (e.g., degree distribution, clustering coefficient, etc.) should
 converge. The actual definition is based on homomorphisms (adjacency
preserving maps) as follows. 
For two simple graphs $F$ and $G$, let $\hom(F,G)$ 
denote the number of homomorphisms from $V(F)$ (the nodes in
$F$) to $V(G)$ (the nodes in $G$). The {\it homomorphism density}
$t(F,G)$ is defined as the probability that a random map from $V(F)$
 to $V(G)$ is a homomorphism, given by
\begin{equation}
t(F,G)= \frac{{\rm hom}(F,G)}{|V(G)|^{|V(F)|}}.
\label{eq:HOMDENSE}
\end{equation}
A sequence $(G_n)$ of graphs is {\it convergent}, if the sequence
$t(F,G_n)$ has a limit for every simple graph $F$.
Losely speaking, this condition can be  interpreted as the convergence of the
probability of finding any given finite sub-graph in the sequence of
networks.

We note that when $G$ has bounded average
degree and $F$ is connected, it is more natural to normalize
the homomorphism density by dividing with $|V(G)|$ instead of
$|V(G)|^{|V(F)|}$. Therefore, we let ${\rm inj}(F,G)$ denote the number 
of injective homomorphisms from
graph $F$ to graph $G$ and define
\begin{equation}
s(F,G)\equiv\frac{{\rm inj}(F,G)}{|V(G)|}.
\end{equation}
(Thus, $s(F,G)$ is the average number of labeled copies of $F$, such
that a specified node of $F$ goes on a specified node of $G$.)

Convergent graph sequences are related to 2d functions in a non-trivial
way \cite{lovasz-szegedy,lovasz_cikk}. 
First of all, we can construct a convergent graph sequence
using a symmetric measurable function, $0\leq W(x,y)\leq 1$, 
defined on the unit square as follows.
For a given network size $N$, we distribute $N$ points independently,
uniformly at random on the $[0,1]$ interval. These points correspond to
 the nodes in the network, and each pair of nodes is linked with
the probability given by $W(x,y)$ at the coordinate of the according points.
In the $N\rightarrow\infty$ limit the obtained graph sequence is converging.
What is even more surprising, it can be proven that we can
 represent any convergent graph sequence by a 2d function, since for
any convergent graph sequence one can find a $W(x,y)$ providing the
 same limiting sub-graph densities.

The average degree of nodes in a random graph generated from a given
$W(x,y)$ using  the construction above can be given simply as
\begin{equation}
\left< d\right>=N\int \int W(x,y)dxdy.
\end{equation}
Thus, in the
$N\rightarrow\infty$ limit the obtained network becomes dense.
In contrast, real networks are usually sparse in the sense that their
 average degree is not expected to grow with increasing size. A solution
to this problem was proposed by Bollob\'as et al., by redefining the
linking probabilities as $W(x,y)/N$, resulting in a network with an
average degree independent of $N$. They showed that depending on the choice
of $W(x,y)$, a wide range of sparse networks can be generated.

Our approach is different from this method in that instead of using
a construction into which we build in the level of complexity from
the beginning, we generate complexity by using tensorial products of
increasing power as $N\rightarrow\infty$. This is a qualitatively
new picture, corresponding to reality to a higher degree
(larger graphs are more complex/inhomogeneous/structured in nature
 than smaller graphs). In addition, we achieve this using a relatively
simple construction.

\subsubsection*{A1.1 Limiting cases of the multifractal model}

A shortcoming of our model is that it can lead to a network in which
the majority of nodes are isolated in the $N\rightarrow\infty$ limit.
However, as we shall see, this effect is negligable for graphs in the
size range of real networks. 

%\subsubsubsection*{A1.Analytical results}
In general, if $W_1(x,y),W_2(x,y),\dots,W_k(x,y)$ is a sequence of 
symmetric measurable 
functions on the unit square (with $0\leq W_k(x,y)\leq 1$ for any $k$),
 let us define $w_k(x)$ as the average linking probability for 
a node at position
 $x$ given by
\begin{equation}
w_k(x)=\int_0^1 W_k(x,y)dy.
\label{eq:w_k_x}
\end{equation}
Similarly, let $\omega_k$ denote the average link probability for the whole 
network, which can be expressed as 
\begin{equation}
\omega_k=\int_0^1w_k(x)dx.
\label{eq:omega_k}
\end{equation}
Let us choose the number of nodes, $N_k$  
associated to $W_k(x,y)$ in such a way that the average degree of nodes
 converges to a constant (non zero) $\left< d\right>$  
for $k\rightarrow\infty$, thus 
\begin{equation}
N_k\omega_k\rightarrow \left< d\right>.
\label{eq:omega_k_lim}
\end{equation} 
 (This means that the number of links is around $\left< d\right>/2$.)
The degree distribution of a node at position $x$ can be given by
 a binoimal distribution as
\begin{equation}
\rho(d,x)=\left(
\begin{matrix}
N_k \\ d
\end{matrix}
\right) w_k(x)^d\left[1-w_k(x)\right]^{N_k-1-d}.
\end{equation}
In the thermodynamic limit this can be approximated by a Poisson-distribution
 written as
\begin{equation}
\rho(d,x)\simeq\frac{\left[N_kw_k(x)\right]^d}{d!}e^{-N_kw_k(x)}.
\end{equation}
The degree distribution of the whole network is obtained by integrating
$\rho(d,x)$, resulting in
\begin{equation}
\rho(d)=\frac{1}{d!}\int_0^1 \left[N_kw_k(x)\right]^de^{-N_kw_k(x)}dx.
\end{equation}
In particular, the probability that a randomly chosen node will be 
isolated (having degree zero) is
\begin{equation}
\rho(d=0)=\int_0^1e^{-N_kw_k(x)}dx.
\end{equation}
From (\ref{eq:omega_k}) and (\ref{eq:omega_k_lim}) it follows that 
the average value of $w_k(x)$ is around $\left< d\right>/N_k$. 
In case $w_k(x)$ is actually independent of $x$, 
then $w_k(x)=\left< d\right>/N_k$, and
\begin{equation}
\rho(d=0)\simeq e^{-\left< d\right>}.
\end{equation}
However, if $w_k(x)$ is such that its typical value is much smaller than
 its average, then typically $e^{-N_kw_k(x)}\simeq 1$ resulting in
\begin{equation}
\rho(d=0)\simeq 1,
\end{equation}
which means that the majority of nodes becomes isolated. The condition
for avoiding this degeneracy can be formulated as
\begin{equation}
\int_0^1e^{-N_kw_k(x)}dx<c,
\label{eq:good}
\end{equation}
where $c<1$ is a constant.

In case of the multifractal graph generator, (or a more general ``tensoring'' 
construction),  the above condition is not fulfilled, unless
$w_k(x)$ is independent of $x$. As mentionned in the main text,
by using a measure preserving bijection between $[0,1]$ and 
$[0,1]^k$, our model can be formulated in a more general form
using the tensorial product $W_k\equiv W^{\bigotimes k}=W\otimes\cdots\otimes W$
defined as
\begin{equation}
W(x_1,\dots,x_k,y_1,\dots,y_k)=W(x_1,y_1)\cdots W(x_k,y_k).
\end{equation}
The marginals (\ref{eq:w_k_x}) in this representation are given by
\begin{eqnarray}
w_k(x_1,\dots,x_k)&=&\int_{[0,1]^k} W(x_1,y_1)\cdots W(x_k,y_k)dy_1\dots dy_k
\nonumber \\ &=&w(x_1)\cdots w(x_k),
\end{eqnarray}
where $w(x)=\int_0^1 W(x,y)dy$. Similarly, (\ref{eq:omega_k}) 
is transformed into
\begin{equation}
\omega_k=\int_{[0,1]^k}w_k(x_1,\dots,x_k)dx_1\dots dx_k=\omega^k,
\end{equation}
where $\omega=\int_0^1 w(x)dx$. Thus, according to (\ref{eq:omega_k_lim}), we
 should choose $N_k\simeq \left< d\right>/\omega^k$. 

Unfortunately, these functions do not satisfy condition
(\ref{eq:good}) unless $w(x)$ is constant. Indeed, if
$(x_1,\dots,x_k,y_1,\dots,y_k)$ is a random point in $[0,1]^k$, then
\begin{equation}
\ln w_k(x_1,\dots,x_k) = \ln w(x_1)+\cdots+ \ln w(x_k) \sim
k\int_0^1\ln w(x)\,dx
\end{equation}
almost surely by the Law of Large Numbers. Let $\alpha=\exp(\int\ln
w(x)\,dx)$, then $\alpha < \omega$ by the Jensen inequality (expect
if $w$ is constant), and the value of $w_k$ is almost always close to
$\alpha^k$, while its average is $\omega^k$. Since
$(\alpha/\omega)^k\to 0$ if $k\to\infty$, this shows that if
\eqref{eq:good} holds, then $w(x)$ is constant.

On the other hand if $w(x)=\omega$ for any $x$, then the
expected degree of the nodes becomes independent from their position and
the degree distribution converges to a Poisson distribution, just like
 in case of an Erd\H os-R\'enyi graph. In this case it 
is also easy to calculate the number of copies of any connected
graph $F$ with $l$ nodes in a graph $G_k$ obtained from $W_k$. 
There are $N_k(N_k-1)\cdots(N_k-l+1)\sim N_k^l$ ways to map $V(F)$ into $[n_k]$
injectively, and for each map, the probability that it is a
homomorphism is $t(F,W_k)=t(F,W)^k$. Hence
\begin{equation}
{\sf E}({\rm inj}(F,G_k))\sim N_k^l t(F,W)^k=\left< d\right>^l
\left(\frac{t(F,W)}{\omega^l}\right)^k.
\end{equation}
Since we have a sparse graph, we want to normalize this by $N_k$; so
the normalized number of copies of $F$ is
\begin{equation}
{\sf E}(s(F,G_k)) = \frac{{\rm hom}(F,G_k)}{N_k} \sim N_k^{l-1} t(F,W)^k=
\left< d\right>^{l-1}
\left(\frac{t(F,W)}{\omega^{l-1}}\right)^k.
\label{eq:E_s}
\end{equation}
For example, the normalized number of triangles is
\begin{equation}
\frac{1}{6} s(K_3,G_k) \sim  \left< d\right>^2
\left(\frac{t(K_3,W)}{\omega^2}\right)^k.
\end{equation}

It is easy to see that if $w(x)=\omega$ is constant, then
\begin{equation}
t(F,W)\leq \omega^{l-1},
\label{eq:tfw}
\end{equation}
where equality holds if and only if $F$ is a tree or $W$ is an
equivalence relation such that there is a partition 
$S_1\cup\dots\cup S_m=[0,1]$ and
\begin{equation}
W(x,y)=
  \begin{cases}
    1, & \text{if $x,y\in S_i$ for some $i$}, \\
    0, & \text{otherwise}.
  \end{cases}
\end{equation}
From (\ref{eq:E_s}) and (\ref{eq:tfw}) we gain
\begin{equation}
{\sf E}(s(F,G_k)) \to
  \begin{cases}
   \left< d\right>^{l-1}, & \text{if $F$ is a tree or $W$ is an
equivalence relation}, \\
   0, & \text{otherwise}.
  \end{cases}
\label{eq:slim}
\end{equation}
Using high concentration inequalities one can prove that this
convergence happens almost surely (not just in expectation).
We see from (\ref{eq:slim}) that the sequence $G_k$ is convergent with
probability $1$ in the Benjamini--Schramm sense.

There are a number of possibilities which one could try to cure 
the degeneracy of the thermodynamic limit of our model shown here,
however these are out of the scope of the present study. We could
 modify the tensoring construction by adding to $W^{\bigoplus k}$ 
a constant $c_k$ tending to $0$ reasonably slowly. Another possibility
is to modify $W^{\bigoplus k}$ to $(W^{\bigoplus k})^{a_k}$, where $a_k\to 0$.

\subsubsection*{A1.2 Numerical studies}
\begin{figure}[h!]
\begin{center}
\centerline{\includegraphics[width=.6\textwidth]{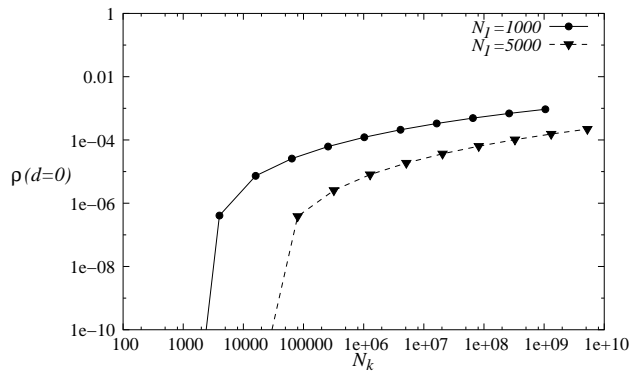}}
\caption{The ratio of isolated nodes in function of the network size for
 graphs obtained from iterating random generating measures from $k=1$ to
 $k=11$, averaged over 1000 samples. The number of nodes at iteration $k=1$
 was chosen to be $N(k=1)=1000$ (circles) and $N(k=1)=5000$ (triangles).
\label{fig:deg_zero}
}
\end{center}
\end{figure}

%\subsubsection{Fraction of isolated nodes}
Next, let us investigate the magnitude of the above effect for graphs in 
the size range of real networks. For this purpose we generated networks
from randomly chosen generating measures (with $m=4$, equal sized
 boxes) iterated from $k=1$ to $k=11$. The number of nodes at $k=1$ was
 set to 1000 and to 5000 respectively, and for $k>1$ it was adjusted
 using eq. () in the main text. (Thus the average degree of the graphs
 remained the same during the iterations.) In Fig.\ref{fig:deg_zero}. we
show the results obtained by averaging over 1000 samples for both 
settings by plotting $\rho(d=0)$ as a function of $N$. Inspite of the 
increasing tendency of the curves, at the last iteration with network sizes
above $10^9$, the fraction of isolated nodes is still very low. 
Thus, the effect of isolated nodes becoming dominant is negligible on
the scale of real world applications.

\begin{figure}[h!]
\begin{center}
\centerline{\includegraphics[width=0.6\textwidth]{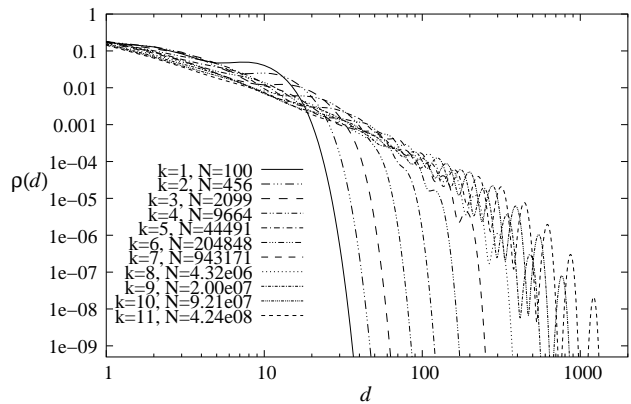}}
\caption{The degree distribution obtained from Eq.(2) in the main text 
for an $m=3$ generating measure iterated from $k=1,N=100$ to $k=11, N=4.24\cdot 10^8$.
\label{fig:oscillate}
}
\end{center}
\end{figure}

%\subsubsection{Absence of convergence in the realistic size range}
In spite of the analytical results for the convergence in the thermodynamic
limit, the degree distribution often shows an oscillatory behavior
in the size range of real networks.
This is shown in Fig.\ref{fig:oscillate} for an $m=3$ generating measure
iterated from $k=1, N=100$ to $k=11, N=4.24\cdot 10^8$. As $k$ is getting larger, the 
more oscillations can be observed in $\rho(d)$ towards the large degrees.

\subsection*{A2 Calculating  statistical distributions}

%\section{Analytic results for the statistics}

A serious advantage of our model is that the statistical properties
characterizing the network topology can be calculated analytically.
In the Appendix of the main text we give a derivation for
 the degree distribution, based on the generating function formalism.
The definition and the
most important properties of the generating functions can be
 summarized as follows.

\subsubsection*{A2.1 The generating functions}
 If a random
variable $\xi$
can take non-negative integer values according to some
probability distribution ${\cal P}(\xi=n)\equiv\rho(n)$,
 then the corresponding generating function
is given by
\begin{equation}
G_{\rho}(x)\equiv\left<x^{\xi}\right>=\sum_{n=0}^{\infty}\rho(n)x^n.
\end{equation}
The generating-function of a properly normalized distribution
is absolute convergent for all $|x|\leq1$ and hence has no singularities
in this region. For $x=1$ it is simply
\begin{equation}
G_{\rho}(1)=\sum_{n=0}^{\infty}\rho(n)=1.
\end{equation}
The original probability distribution and its moments can be
obtained from the generating-function as
\begin{eqnarray}
\rho(n)&=&\frac{1}{n!}\left.\frac{d^nG_{\rho}(x)}{dx^n}\right|_{x=0},
\label{eq:vissza}\\
\left<\xi^l\right>&=&\sum_{n=0}^{\infty}n^l\rho(n)=\left[\left(x\frac{d}{dx}
\right)^lG_{\rho}(x)\right]_{x=1}.\label{eq:moment}
\end{eqnarray}
And finally, if $\eta=\xi_1+\xi_2+...+\xi_l$, where $\xi_1,\xi_2,...,\xi_l$ are
independent random variables (with non-negative integer values),
then the generating function corresponding to
${\cal P}(\eta=n)\equiv\sigma(n)$ is given by
\begin{equation}
G_{\sigma}(x)=\left<x^{\eta}\right>=\left<\prod_{i=1}^{I}x^{\xi_i}\right>=
\prod_{i=1}^{I}\left<x^{\xi_i}\right>=G_{\rho_1}(x)G_{\rho_2}(x)\cdots G_{\rho_l}(x).
\label{eq:powers}
\end{equation}

\subsubsection*{A2.2 The degree distribution}
The degree distribution of the nodes falling in row $i$ of the link
 probability measure, $\rho_i^{(k)}(d)$, can be calculated as follows.
In our construction we draw links for a node in row $i$ pointing to
nodes in row $j$ altogether $n_j(k)$ times with a probability $p_{ij}(k)$, 
where $n_j(k)$ is the number of nodes in row $j$, given by
\begin{equation}
n_j(k)=Nl_{j}(k).
\end{equation}
Since the link draws are independent, the distribution of the number of
 links from a node in
row $i$ to nodes in row $j$ is binomial. This can be
approximated by a Poisson-distribution when $n_j$ is sufficiently large as
\begin{eqnarray}
\rho^{(k)}_{ij}(d)&=&%{n_j(k),d}
\left(
\begin{matrix}
n_j(k) \\ d
\end{matrix}
\right)
[p_{ij}(k)]^d[1-p_{ij}(k)]^{n_j(k)-d} \simeq
\nonumber \\ & &\frac{\left< d_{ij}(k)\right>^{d}}{d!}e^{-\left< d_{ij}(k)\right>},
\end{eqnarray}
where $\left< d_{ij}\right>$ denotes the average number
of links from a
 node in row $i$ to nodes in row $j$ given by
\begin{equation}
\left<d_{ij}\right>=n_j(k)p_{ij}(k).
\label{eq:d_ij}
\end{equation}
 The degree of a node in row
 $i$ is given by the sum over the links towards the other rows as
\begin{equation}
d_i(k)=\sum_{j}d_{ij}(k).
\end{equation} 
Therefore, the generating function
 of $\rho_{i}^{(k)}(d)$ is the product of the generating functions of
the $\rho^{(k)}_{ij}(d)$ distributions:
\begin{equation}
G_i^{(k)}(x)=\prod_j G_{ij}^{(k)}(x),
\label{eq:gen_prod}
\end{equation}
where $G_{ij}^{(k)}(x)$ is defined as
\begin{eqnarray}
G_{ij}^{(k)}(x)&=&\sum_{d}^{\infty}\rho_{ij}^{(k)}(d)x^{d}=\sum_{d=0}^{\infty}\frac{\left< d_{ij}(k)\right>^{d}}{d!}e^{-\left< d_{ij}(k)\right>}x^d
= \nonumber \\ & & e^{\left< d_{ij}(k)\right>(x-1)}.
\label{eq:G_ij}
\end{eqnarray}
(A summary of the most important properties of the generating functions is
 given in the Supporting Information). By substituting (\ref{eq:G_ij}) into
(\ref{eq:gen_prod}) we arrive to
\begin{equation}
G_i^{(k)}(x)=\prod_je^{\left< d_{ij}(k)\right>(x-1)}=e^{(x-1)\sum_j\left< d_{ij}(k)\right>}=e^{(x-1)\left< d_i(k)\right>},
\label{eq:gen_final}
\end{equation}
where we used that due to the independence of the links,
the expected degree of a node in row $i$ can be expressed as
\begin{equation}
\left< d_i\right>=\sum_j\left< d_{ij}\right>.
\label{eq:d_i}
\end{equation}
An alternative form for $\left< d_i\right>$ can be obtained by 
substituting (\ref{eq:d_ij}) into the equation above, yielding
\begin{equation}
\left< d_i\right>=N\sum_{j}p_{ij}(k)l_{j}(k).
\end{equation}
The degree distribution of the nodes falling into row $i$ can be
obtained by transforming back
the generating function in (\ref{eq:gen_final}), resulting in
\begin{equation}
\rho_i^{(k)}(d)=\frac{\left< d_i(k)\right>^{d}}{d!}e^{-\left< d_i(k)\right>}.
\end{equation}

\subsubsection*{A2.3 The clustering coefficient}
Similarly to the degree distribution, the clustering coefficient of nodes
falling into the same row of the link probability measure is expected to
be the same. The
clustering coefficient of a node in row $i$ can be obtained by calculating
the number of triangles containing the node, divided by the
number of link pairs originating from the node. Since the triangles are equivalent
to link pairs originating from the node having their other end connected by
a third link, the expected clustering coefficient of a node in row $i$ can
be given as
%\begin{figure*}
\begin{equation}
\left< C_i(k)\right>=\frac{\frac{1}{2}\sum_{j=1}^{m^k} [l_j(k)]^2
\left[p_{ij}(k)\right]^2p_{jj}(k)+\sum_{j=1}^{m^k}\sum_{q=j+1}^{m^k}l_j(k)l_q(k)p_{ij}(k)p_{iq}(k)p_{jq}(k)}{\frac{1}{2}\sum_{j=1}^{m^k} [l_j(k)]^2\left[p_{ij}(k)\right]^2+\sum_{j=1}^{m^k}\sum_{p=j+1}^{m^k}l_j(k)l_q(k)p_{ij}(k)p_{iq}(k)}.
\label{eq:clust_coeff}
\end{equation}
%\end{figure*}
The first term in both the numerator and the denominator corresponds to the
link pairs for which the other end of the links point to the same row $j$,
whereas the second terms give the contribution from link pairs connecting our
node in row $i$ to distinct rows $j$ and $q$.
%In Fig.\ref{fig:clust}. we show the empirical $\left< C_i \right>$ together
% with the result from (\ref{eq:clust_coeff}) for a randomly chosen
%generating measure at $m=3$, $k=4$ and $N=10000$. Although the clustering
%coefficient is varying abruptly from one row to the other, the two
% results are in very good agreement with each other.

\subsubsection*{A2.4 The average nearest neighbors degree}
Finally, we mention that the degree correlations can be calculated from
$p_{ij}(k)$ (and $l_i(k)$) as well. Here we derive
the expression for the average nearest neighbors degree, $d_{NN}$,
as a function of the node-degree.  This is one of the most
 simplest quantity characterizing the degree correlations: an increasing curve
corresponds to an assortative network, whereas a decreasing one signals
disassortative behavior.
The average
 degree of the neighbors of a node from row $i$ can be given as
 \begin{equation}
d_{NN,i}^{(k)}=\frac{\sum_{j=1}^{m^k} \widehat{p}_{ij}(k)l_j(k)\left< d_j(k)\right>}{\sum_{j=1}^{m^k} p_{ij}(k)l_j(k)}.
\label{eq:dNN_i}
 \end{equation}
The average
 degree of the neighbors of a node with degree $d$ can be given as a sum
 over the possible $d_{NN,i}^{(k)}$, multiplied by the conditional probability
 $p^{(k)}(i|d)$ that the node is from row $i$, given that its degree is $d$:
\begin{equation}
d_{NN}^{(k)}(d)=\sum_{i=1}^{m^k}p^{(k)}(i|d)d_{NN,i}^{(k)}.
\label{eq:dNN_alap}
\end{equation}
These conditional probabilities can be obtained as follows. The number
of nodes from row $i$ with degree $d$ is $n_i(k)\rho_i^{(k)}(d)$, whereas the total
 number of nodes with degree $d$ is $n\rho^{(k)}(d)$.
The probability that a node is from row $i$ given that its degree
 is $d$ is the ratio of these two:
\begin{equation}
p^{(k)}(i|d)=\frac{n_i(k)\rho_i^{(k)}(d)}{n\rho^{(k)}(d)}=\frac{l_i(k)\rho_i^{(k)}(d)}
{\rho^{(k)}(d)}.
\label{eq:cond_prob}
\end{equation}
By substituting (\ref{eq:cond_prob}) and (\ref{eq:dNN_i}) into
(\ref{eq:dNN_alap}) we get
\begin{equation}
d_{NN}^{(k)}(d)=\frac{1}{\rho^{(k)}(d)}\sum_{i=1}^{m^k}l_i(k)\rho_i^{(k)}(d)\frac{\sum_{j=1}^{m^k} p_{ij}(k)l_j(k)\left< d_j(k)\right>}{\sum_{j=1}^{m^k} p_{ij}(k)l_j(k)}.
\label{eq:ANND}
\end{equation}

%The distributions associated with the clustering coefficients and
%the degree correlations can be derived analogously as described in
%the Supporting Information.


\begin{thebibliography}{10}


\bibitem{albert-revmod}
Albert~R, Barab{\'a}si~A-L (2002)
Statistical mechanics of complex networks.
{\em Rev.\ Mod.\ Phys. 74}: 47--97.

\bibitem{dm-book}
Mendes~JFF, Dorogovtsev~SN (2003)
{\em Evolution of Networks: From Biological Nets to the Internet and WWW}
(Oxford University Press, Oxford).

\bibitem{newman-book}
Newman~MEJ, Barab\'asi~A-L, Watts~DJ (2006) {\em The Structure and Dynamics of Networks} (Princeton University Press, Princeton).

\bibitem{watts-strogatz}
Watts~DJ, Strogatz~SH (1998)
Collective dynamics of 'small-world' networks.
{\em Nature 393}: 440--442.

\bibitem{F3}
%Faloutsos M \etal
Faloutsos~M, Faloutsos~P, Faloutsos~C (1999) On Power-Law Relationships of the Internet Topology. {\em Comput.\ Commun.\ Rev.\ 29}: 251--262.

\bibitem{barabasi-albert}
Barab{\'a}si~A-L, Albert~R (1999)
Emergence of scaling in random networks.
{\em Science 286}: 509--512.

\bibitem{gn-pnas}
Girvan~M, Newman~MEJ (2002) 
Community structure in social and biological networks.
{\em Proc.\ Natl.\ Acad.\ Sci.\ USA 99}: 7821--7826.

\bibitem{our-nature}
Palla~G, Der\'enyi~I, Farkas~I, Vicsek~T (2005)
Uncovering the overlapping community structure of complex networks in 
nature and society. 
{\em Nature 435}: 814--818.

\bibitem{fortunato-new}
Lancichinetti~A, Fortunato~S, Kert\'esz~J (2009) {\em Detecting the overlapping and hierarchical community structure of complex networks}
New Journal of Physics 11: 033015.

\bibitem{laci-disease}
Goh~K-I, Cusick~ME, Valle~D, Childs~B, Vidal~M, Barab\'asi~A-L (2007)
The human disease network.
{\em Proc.\ Natl.\ Acad.\ Sci.\ USA  104}: 8685--8690.

\bibitem{drawing_contest}
Dogrusoz~U, Duncan~CA, Gutwenger~C, Sander~G (2009)
Graph Drawing Contest Report.
{\em Lect.\ Notes in Comp.\ Sci.\ 5417}: 453--458.

\bibitem{lovasz-szegedy}
Lov\'asz~L, Szegedy~B (2006) Limits of dense graph sequences, 
{\em J.\ Comb.\ Theory B 96}: 933--957.

\bibitem{lovasz_cikk}
Borgs~C, Chayes~J, Lov\'asz~L, S\'os~VT, Vesztergombi~K (2008) 
Convergent Sequences of Dense Graphs I: Subgraph Frequencies, 
Metric Properties and Testing.
{\em  Advances in Math.\ 219}: 1801--1851.

\bibitem{Bollobas-cikk}
Bollob\'as~B, Janson~S, Riordan~O (2007) 
The phase transition in inhomogeneous random graphs.
{\em Random Structures \& Algorithms 31}:3--122.

\bibitem{Bollobas_chapter}
Bollob\'as~B, Riordan~O (2009) 
Random Graphs and Branching Process. 
in {\em Handbook of Large-scale Random Networks}, 
eds Bollob\'as~B, Kozma~R, Mikl\'os~D 
(Springer, Berlin) pp 15--115.

\bibitem{dk-series}
Mahadevan~P, Krioukov~D, Fall~K, Vahdat~A (2006)
Systematic topology analysis and generation using degree correlations.
{\em ACM SIGCOMM Computer Communication Review 36}: 135--146.

\bibitem{Frank-exp}
Frank~O, Strauss~D (1986)
Markov graphs.
{\em Journal of the American Statistical Association 81}: 832--842.

\bibitem{Wasserman-exp}
Wasserman~S, Pattison~PE (1996)
Logit models and logistic regressions for  social networks: I. An introduction
 to Markov graphs and p*.
{\em Psychometrika 61}: 401--425.

\bibitem{snijders-exp}
Robins~G, Snijders~T, Wang~P, Handcock~M, Pattison~P (2007)
Recent developments in exponential random graph (p*) models for social 
networks.
{\em Social Networks 29}: 192--215.

\bibitem{Tamas_fract}
Barab\'ssi~A-L, Ravasz~E, Vicsek~T (2001) Deterministic scale-free networks. {\em Physica A 299}: 559--564.

\bibitem{Havlin_fract}
Song~CM, Havlin~S, Makse~HA (2005) Self-similarity of complex networks.
 {\em Nature 433}: 392--395.


\bibitem{Newman_hier}
Clauset~A, Moore~C, Newman~MEJ (2008) Hierarchical structure and the prediction
of missing links in networks. {\em Nature 453}: 98--101.

\bibitem{Avetisov}
Avetisov~VA, Chertovich~AV, Nechaev~SK, Vasilyev~OA (2009)
On scale-free and poly-scale behaviors of random hierarchical network.
{\em  J. Stat. Mech. 2009}: P07008.


\bibitem{Parisi_matrix}
Mezard~G, Parisi~G, Virasoro~M (1987) {\em Spin glass theory and beyond} (World
Scientific, Singapore).

\bibitem{E-R}
Erd\"os~P, R\'enyi~A (1960) On the evolution of random graphs. {\em Publ.\ Math.\ Inst.\ Hung.\ Acad.\ Sci.\ 5}: 17--61.


\bibitem{Leskovec_elso}
Leskovec~J, Chakrabarti~D, Kleinberg~J, Faloutsos~C (2005)
Realistic, Mathematically Tractable Graph Generation and Evolution, 
Using Kronecker Multiplication.
{\em European Conference on Principles and Practice of Knowledge Discovery in Databases:} 133-145.

\bibitem{Leskovec_two}
Leskovec~J, Faloutsos~C (2007)
Scalable Modeling of Real Graphs using Kronecker Multiplication.
{\em Proceedings of the 24th International Conference on Machine Learning}: 
497--504.

\bibitem{Nepusz_chapter}
Nepusz~T, N\'egyessy~L., Tusn\'ady~G., Bazs\'o~F (2008) Reconstructing cortical
networks: case of directed graphs with high level of reciprocity.
in {\em, Handbook of Large-scale Random Networks},
eds Bollob\'as~B, Kozma~R, Mikl\'os~D
(Springer, Berlin) pp 325--368.

\bibitem{Nepusz_cikk}
Nepusz~T, Bazs\'o~F (2007) Likelihood-based clustering of directed graphs {\em IEEE Proceedings of the 3rd International Symposium on Computational Intelligence and Intelligent Informatics}: 189--194.


\bibitem{Szemeredy}
Szemer\'edy~E (1978) Regular partitions of graphs. in {\em Probl\'emes combinatoires et th\'eories des graphes.} (Centre National de la Recherche Scientifique)
 pp 399-401.



\bibitem{Molloy_Reed_95}
Molloy~M, Reed~B (1995)
A critical point for random graphs with a given degree sequence.
{\em Random Structures \& Algorithms 6}: 161-179.


\bibitem{Molloy_Reed_98}
Molloy~M, Reed~B  (1998)
The size of the giant component of a random graph with a given degree sequence.
{\em Combinatorica Probability and Computing 7}: 295--305.

\bibitem{Chung_lu}
Chung~F, Lu~L (2002)
The average distances in random graphs with given expected degrees.
{\em Proc.\ Natl.\ Acad.\ Sci.\ USA 99}: 15879--15882.

\bibitem{Boguna}
Serrano~M{\'A}, Bogu{\~n}{\'a}~M (2005)
Tuning clustering in random networks with arbitrary degree distributions.
{\em Phys.\ Rev.\ E 72}: 036133.

\bibitem{Metropolis}
Metropolis~N, Rosenbluth~AW, Rosenbluth~MN, Teller~AH, Teller~E (1953)
Equations of State Calculations by Fast Computing Machines.
{\em J.\ Chem.\ Phys.\ 21}: 1087-1092.

\end{thebibliography}
\end{document}